\documentclass{article}
\usepackage[dvips]{graphics}
\usepackage{array,hhline,amssymb,amsthm,color,amsfonts}
\def\L{{\mathfrak{L}}}
\def\O{{\mathcal{O}}}  \def\o{{\mathbf{o}}}
  \def\k{{\overline{\o}}}
\def\U{\mathfrak{O}}   
  \def\C{\mathbb{C}}    
\def\d12{\raise0.4ex\hbox{\scriptsize 1}\!/\!\lower0.4ex\hbox{\scriptsize 2}}
\def\tr{{\rm tr}}

\def\bux#1#2{\buildrel{*}\over{#1}_{#2}}

\title{The Riemann-Cartan space in the O-theory}
\date{}
\author{V. Yu. Dorofeev\thanks{Dep. of Math., SPb SUEF,
Sadovaya 21, 191023, St.Petersburg,
Russia, E-mail: friedlab@mail.ru}\\
Friedmann Laboratory for Theoretical Physics}
\begin{document}
\maketitle
\begin{abstract}
Nonrelativistic equation of particle with a spin for the Lagrangian
on a nonassociative algebra is obtained. It is shown that in this
model arises Riemann-Cartan space. In the case of central symmetry
in addition to the pseudo-curvature appears torsion as pseudovector
that interacts with the spin of the particle. An estimation of the
influence of torsion on the strength of gravitational attraction in
the central gravitational field is given.
\end{abstract}
\section{Introduction}
In the paper \cite{Dor1} the author proposes a generalization of the
Lagrangian theory of Weinberg-Salam on the nonassociative algebra.
In the paper \cite{Dor2} the author considered some physical
consequences of the nonassociative Lagrangian. In particular, it was
found that the proposed model is consistent with geometric solutions
of general relativity types of Schwardzchild and Friedmann metric,
and Einstein's equations of the form
$$T_{\mu\nu}=\lambda R_{\mu\nu}$$

Octonions as an algebraic structure for the construction of physical
theories are used by a number of authors (\cite{Nest}, \cite{Gob},
etc.). In order to identity the theory, which is based on the
Lagrangian, developed in \cite{Dor1}, the name of O-theory is
proposed.

This work is a direct continuation of the research started in
\cite{Dor1} and \cite{Dor2}.

\section{Concepts of the O-theory}
Recall some fundamental aspects of O-theory \cite{Dor1}.

Compensating fields ${\bf A_\mu}(x),x\in M_4$ in the Minkowski space
$M_4$ of nonassociative Lagrangian $\L_o$ have the form:
\begin{equation}\label{kplo}
{\bf A_\mu}(x)=A_\mu^i(x)\Sigma^i,i=0,1,2,\dots,7
\end{equation}
where $\Sigma^i$ are generators of the algebra $\O$ and $A_\mu^i(x)$
are vector fields.

Nonassociative algebra is represented by the matrices ($i=1,2,3$)
\begin{equation}\label{Zornpr}
\matrix{\Sigma^0=\left(\matrix{\sigma^0&0\cr0&\sigma^0}\right)&\Sigma^i=
\left(\matrix{0&-i\sigma^i\cr i\sigma^i&0}\right)\cr
\Sigma^4=\left(\matrix{-\sigma^0&0\cr0&\sigma^0}\right)&
\Sigma^{4+i}=\left(\matrix{0&-\sigma^i\cr -\sigma^i&0}\right)}
\end{equation}
where $\sigma^0$ is an identity matrix and $\sigma^i,i=1,2,3$ are
Pauli matrices:
\begin{equation}\label{matDir}
\sigma^1=\left(\matrix{0&1\cr1&0}\right),\quad
\sigma^2=\left(\matrix{0&-i\cr i&0}\right),\quad
\sigma^3=\left(\matrix{1&0\cr0&-1}\right),\quad
\sigma^0=\left(\matrix{1&0\cr0&1}\right)
\end{equation}
with a special law of multiplication
$$\o*\o'=\left(\matrix{\lambda I&A\cr B&\xi I}\right)*
\left(\matrix{\lambda' I&A'\cr B'&\xi' I}\right)=$$
\begin{equation}\label{Dab}
=\left(\matrix{(\lambda\lambda'+\frac12\tr(AB'))I\hfill&\lambda
A'+\xi' A+\frac i2[B,B']\hfill\cr\lambda'B+\xi B'-\frac
i2[A,A']&(\xi\xi'+\frac12\tr(BA'))I\hfill}\right)
\end{equation}

In the last formula we introduced the matrix $A,A',B,B$ of dimension
$(2\times2)$, identity matrix $I=\sigma^0$ and complex numbers
$\lambda,\xi,\lambda',\xi'$.

States space is based on the expanded space $\U$ of the algebra
$\O_\C$. The basis of this space is formed by matrixes
\begin{equation}\label{dmo}
f^a=\Sigma^a,\quad a=0,1,\dots,7,f^8=\left(\matrix{0&I\cr
I&0}\right),\quad f^9=\left(\matrix{0& iI\cr-iI&0}\right)
\end{equation}
keeping product from (\ref{Dab}).

By means of multiplication (\ref {Dab}) we introduce convolution:
\begin{equation}\label{skpro}
(\k_1,\k_2)=\frac12\tr(\k_1^+*\k_2)=\frac12\tr(\left(\matrix{\lambda_1^*I&B_1^+\cr A_1^+&\xi_1^*I}\right)*
\left(\matrix{\lambda_2I&A_2\cr B_2&\xi_2I}\right))$$
$$=\lambda_1^*\lambda_2+\xi_1^*\xi_2+\frac12\tr(B_1^+B_2))
+\frac12\tr(A_1^+A_2)
\end{equation}
where $\U\times\U$ maps to $\C$.

\section{Nonassociative Lagrangian}
In the work \cite {Dor1} the proposed Lagrangian, generalising
Weinberg-Salam Lagrangian (Standart Theory=ST) on the nonassociative
algebra, has a following form ($a,b=0,1,\dots,7,k=1,2,\dots,7$):
$$\L_{oct.}=\L_f+(\partial_\mu\bux\Psi\varphi-
\frac i2q^aA_\mu^{a}\bux\Psi\varphi*\Sigma^{a})
*(\partial^\mu\Psi_\varphi+
\frac i2q^{b}A^{\mu(b)} \Sigma^{b}*\Psi_\varphi)$$
$$+\frac i2\overline L*\gamma_\mu(\overrightarrow\partial^\mu
L+\frac i2c_Lq^kA^{\mu(k)}\Sigma^k*L+\frac
i2c_{L_0}q^0A^{\mu(0)}L)$$
$$-\frac i2\overline L*\gamma_\mu(\overleftarrow\partial^\mu L-
\frac i2c_Lq^kA^{\mu(k)}\Sigma^{k}*L-\frac
i2c_{L_0}q^0A^{\mu(0)}L)$$
$$+\frac i2\overline R\gamma_\mu(\overrightarrow\partial^\mu R+
iq^0A^{\mu0}R)-\frac i2\overline
R\gamma_\mu(\overleftarrow\partial^\mu R-iq^0A^{\mu0}R)$$
\begin{equation}\label{vst}
-\tilde h\overline L*\Psi_\varphi R-\tilde h\overline
R\bux\Psi\varphi*L)+m^2||\Psi_\varphi||^2-\frac f4||\Psi_\varphi||^4
\end{equation}

Denote Lagrangian of free fields $\L_f$
\begin{equation}\label{svs0}
\L_f=-\frac14F^0_{\mu\nu}F^{\mu\nu(0)}-\frac14\tr(F_{\mu\nu}^k*F^{\mu\nu(k)})
$$ $$+\frac1{16}f^{ijkl}(A_\mu^iA_\nu^j-A_\nu^j
A_\mu^i)(A^{\mu(k)}A^{\nu(l)}-A^{\nu(k)}A^{\mu(l)})
\end{equation}
\begin{equation}\label{l04}
f^{ijkl}=q^{ij}q^{kl}\tr(\Sigma^i*\Sigma^j*\Sigma^k*\Sigma^l)
\end{equation}
$$F^0_{\mu\nu}=\partial_\mu A^0_\nu-\partial_\nu A^0_\mu,\qquad
F_{\mu\nu}^k=\partial_\mu A_\nu^k-\partial_\nu A_\mu^k-\varepsilon^{ijk}q^{ij}(A_\mu^iA_\nu^j-A_\nu^iA_\mu^j)$$

The values $f^{ijkl}, i, j, k, l=1,\dots,7$ reflect nonassociative
character of $\L_f$.

Introduce the left and the right components of spinors
\begin{equation}\label{lpm}
\frac12(1+\gamma^5)\Psi=L,\qquad\frac12(1-\gamma^5)\Psi=R,\qquad\gamma^5=i\gamma^0\gamma^1\gamma^2\gamma^3
\end{equation}

$\Psi,\Psi_\varphi$ are vectors of states from $\U$ which elements
are bispinors; $q^a,c_L,c_{L_0}$ are some numbers caused by
normalization; $\gamma^\mu$ are Dirac's matrices
($\mu=0,1,2,3;i=1,2,3$):
\begin{equation}\label{matDir2}
\gamma^0=\left(\matrix{I&0\cr0&-I}\right),\qquad
\gamma^i=\left(\matrix{0&\sigma^i\cr-\sigma^i&0}\right),\qquad \gamma^5=\left(\matrix{0&I\cr I&0}\right)
\end{equation}
where different-case Greek indexes specify summation by the metric
tensor of Minkowski space $\eta_{\mu\nu}$ with the signature
$(1,-1,$ $-1,-1)$, and the indexes of the same case specify the
simple summation.

Furthermore, in \cite{Dor1} in the basic vacuum state
$$\Psi_\varphi=\Psi_0=\frac{m}{\sqrt{2f}}\left(\matrix{0&i\sigma^3\cr0&I}\right)$$
the general expression of nonassociative Lagrangian for lepton
sector is derived
$$\L_o=\L_f+
\frac{q^{(k)2}m^2}{2f}A_\mu^{k}A^{\mu(k)}+\frac{o^{ij}m^2}{2f}A_\mu^{i}A^{\mu(j)}+
\frac{g^{(1)2}m^2}{2f}B_\mu B^\mu
-\frac{gg^{(1)}m^2}fA_\mu^3B^\mu$$
$$+\frac{g^{(1)}}2\overline\nu_L\gamma_\mu B^\mu\nu_L+
\frac{g^{(1)}}2\overline e_L\gamma_\mu B^\mu e_L+\frac g2\overline
e_L\gamma_\mu A^{\mu3}e_L-\frac g2\overline\nu_L\gamma_\mu
A^{\mu3}\nu_L$$
$$-\frac g2\overline\nu_L\gamma_\mu e_L(A^{\mu1}-iA^{\mu2})-
\frac g2\overline e_L\gamma_\mu\nu_L(A^{\mu1}+iA^{\mu2})$$
$$+\frac i2(\overline e_L\gamma_\mu\partial^\mu e_L-\partial^\mu\overline
e_L\gamma_\mu e_L)+\frac
i2(\overline\nu_L\gamma_\mu\partial^\mu\nu_L-
\partial^\mu\overline\nu_L\gamma_\mu\nu_L)+\frac{m^4}f$$
$$+\frac i2(\overline e_R\gamma_\mu\partial^\mu
e_R-\partial^\mu\overline e_R\gamma_\mu e_R)+g^{(1)}\overline
e_R\gamma_\mu B^\mu e_R-\frac{\sqrt2hm}{\sqrt f}(\overline
e_Le_R+\overline e_Re_L)$$
$$-q^4A^{\mu(4)}
(\kappa_1\overline\nu_L\gamma_\mu\nu_L- \kappa_2\overline
e_L\gamma_\mu e_L)-
\frac32q^6A^{\mu(6)}\overline e_L\gamma_\mu e_L$$
\begin{equation}\label{plna}
-\frac54(q^6A^{\mu(6)}+iq^5A^{\mu(5)})\overline\nu_L\gamma_\mu
e_L-\frac54(q^6A^{\mu(6)}-iq^5A^{\mu(5)})\overline e_L\gamma_\mu
\nu_L
\end{equation}
with the following denotes $ic_2^{ij}A^{ij}q^iq^j=o^{ij}$.

In \cite{Dor2} it is shown that in case of a certain relation
between the senior fields $A_\mu^i, i=4,5,6,7$ and charge constants
$q^{ij}, i, j=4,5,6,7$ it follows from Lagrangian (\ref {plna}) that
a gravitational field consists of two oppositely charged massless
vector bosons $D_\mu $ and $D^*_\mu $. Thus the Dirac equation for
lepton sector, with the exception of electromagnetic interaction,
looks like:
\begin{equation}\label{urdo}
(i\gamma^\mu(\partial_\mu-iq_DD^\mu_L)-m_e)e(x)=0
\end{equation}
where
\begin{equation}\label{ovd}
q_DD^\mu=-\frac32q^6A^{\mu(6)}-i\frac32q^5A^{\mu(5)}
\end{equation}
and $D_L^\mu=D^\mu\frac12(1+\gamma^5)$.

\section{The Dirac equation in Riemannian space}
Let $M $ be a pseudoriemannian manifold where it is possible to
introduce coordinates $(x^0,x^1,x^2,x^3)$ and metric
\begin{equation}\label{pm}
ds^2=g_{\mu\nu}dx^\mu dx^\nu
\end{equation}

Let $\delta A^\mu=\Gamma^\mu_{\nu\lambda} A^\nu dx^\lambda$
\cite{Eisenhart}, where $\Gamma^\mu_{\nu\lambda}$
\begin{equation}\label{sk} 
\Gamma^\mu_{\nu\lambda}=\frac12g^{\mu\kappa}(g_{\mu\kappa,\nu}+
g_{\nu\kappa,\lambda}-g_{\lambda\nu,\kappa})
\end{equation}

Then for a covariant derivative
\begin{equation}\label{kpr}
A^\mu_{;\nu}=A^\mu_{,\nu}+\Gamma^\mu_{\nu\lambda}A^\lambda
\end{equation}
the transformation to the new coordinates is equivalent to the
transformation of a vector in Riemannian space.

Christoffel symbols don't form a tensor. The combination of
Christoffel symbols is a Riemann tensor which we define in
accordance to \cite{Eisenhart}
\begin{equation}\label{tkr}
R^\tau_{\mu\nu\lambda}=\Gamma^\tau_{\mu\lambda,\nu}-
\Gamma^\tau_{\mu\nu,\lambda}+\Gamma^\tau_{\sigma\nu}\Gamma^\sigma_{\mu\lambda}-
\Gamma^\tau_{\sigma\lambda}\Gamma^\sigma_{\mu\nu}.
\end{equation}
satisfies to the tensor law of transformation.

The quadratic form (\ref{pm}) in a vicinity of some point can be
reduced to a diagonal form
\begin{equation}\label{ivsok2}
ds^2=H^{(0)2}dx^{(0)2}-
H^{(1)2}dx^{(1)2}-H^{(2)2}dx^{(2)2}-H^{(3)2}dx^{(3)2}
\end{equation}

By associating the diagonal coordinates with the coordinates of the
physical space-time, we believe that in some coordinates the metric
is given by the diagonal form of Minkowski space $M_4$:
\begin{equation}\label{mpm}
ds^2=c^2dt^2-dx^2-dy^2-dz^2=\eta_{ab}dx^adx^b
\end{equation}

(In this section we will distinguish Greek and Latin indices,
assuming that the Greek indices refer to the pseudoriemannian space,
while Latin refer to Euclidean one.)

Suppose that at each point of a pseudoriemannian space $M$ there is
defined a stratification, which is the tangent Minkowski space $M_4$
with the metric (\ref{mpm}).

Thus the pseudoriemannian coordinates of the physical space-time and
the coordinates of the tangent space are given by the formulas:
\begin{equation}\label{frm}
H^{(0)}dx^{(0)}=cdt,H^{(1)}dx^{(1)}=dx,H^{(2)}dx^{(2)}=dy,H^{(3)}dx^{(3)}=dz
\end{equation}

On the other hand, in each point of pseudoriemannian space $M$ there
can be introduced tetrads $h^a_\mu$, connecting the metric of
pseudoriemannian space to the metric of Minkowski space:
\begin{equation}\label{v010}
h^b_\mu h^\mu_\nu=\delta^b_a,\qquad h^{\mu(a)}h^\nu_a=g^{\mu\nu}
\end{equation}

Any vector pseudoriemannian spaces $A_\mu$ can be presented
componentwise in space Minkovsky $M_4$ by means of tetrads $h_\mu^a$
as $A^a=A^\mu h^a_\mu$, therefore (\ref{kpr})
\begin{equation}\label{v1}
\delta A^\mu=\delta(A^ah^\mu_a)=\delta A^ah^\mu_a+A^a\delta h^\mu_a=\delta A^ah^\mu_a+A^ah^\mu_{a,\nu}\delta x^\nu=\Gamma^\mu_{\nu\lambda}A^\nu\delta x^\lambda \end{equation}

With the assumption $h^b_\mu h^\mu_{a,\nu}+h^b_{\mu,\nu} h^\mu_a=0$,
find
\begin{equation}\label{t2}
\delta A^b=\gamma^b_{ac}A^a\delta x^c,\qquad\gamma^b_{ac}=
h^b_{\mu;\nu}h^\mu_ah^\nu_c
\end{equation}
where $\gamma^b_{ac}$ are the factors of rotation of Ricci.

In the work of Fock-Ivanenko \cite{FokIvanenko} it is shown that the
free Dirac equation
\begin{equation}\label{urdir}
(i\gamma^a\partial_a-m)e(x)=0
\end{equation}
rewritten in $M_4$ as in tangent layer of a pseudoriemannian
manifold looks like
\begin{equation}\label{urdirr}
(i\gamma^a(\partial_a-\Gamma_a)-m)e(x)=0
\end{equation}
where
\begin{equation}\label{v013}
\Gamma_a=-\frac12\gamma_{abc}\sigma_{bc},\qquad \sigma^{ab}=\frac14[\gamma^a,\gamma^b],\qquad a,b=0,1,2,3
\end{equation}

Dirac matrices $\gamma^a,a=0,\dots,3$ satisfy to the following
multiplication law with matrix  $\sigma^{ab}$:
\begin{equation}\label{v012d}
\gamma^a\sigma^{bc}=\frac14\gamma^a[\gamma^b,\gamma^c]=\frac12\eta^{ab}\gamma^c-\frac12\eta^{ac}\gamma^b-\frac i2\varepsilon^{dabc}\gamma^5\gamma_d
\end{equation}

Hence
\begin{equation}\label{ivsok20}
-\gamma^a\Gamma_a=
\frac14h^\mu_ah^\nu_bh_{(c)\nu;\mu}(\frac12\eta^{ab}\gamma^c-\frac12\eta^{ac}\gamma^b-\frac i2\varepsilon^{dabc}\gamma^5\gamma_d)
\end{equation}

Let the metric of pseudoriemannian space have the form
(\ref{ivsok2}) therefore for nonequal values of $a,b,c$ we have
\begin{equation}\label{ivsok201}
\Gamma^\lambda_{\mu\nu}h_{c\lambda}h^\mu_bh^\nu_c=\frac12g^{\mu\kappa}(g_{\mu\kappa,\nu}+
g_{\nu\kappa,\lambda}-g_{\lambda\nu,\kappa})h_{c\lambda}h^\mu_bh^\nu_c=0
\end{equation}
hence
\begin{equation}\label{ivsok202}
-\gamma^a\Gamma_a=\frac14h^\mu_ah^{\nu(a)}h_{(c)\nu;\mu}\gamma^c-\frac14
h^{\mu(a)}h^\nu_bh_{(a)\nu;\mu}\gamma^b$$
$$=\frac14h^\mu_{c;\mu}\gamma^c+\frac14h^{\mu(a)}h^\nu_{b;\mu}h_{(a)\nu}\gamma^b=\frac12h^\mu_{c;\mu}\gamma^c
\end{equation}
and the Dirac equation forms \cite{SokolovIvanenko}:
\begin{equation}\label{ivsok3}
(\gamma^a(H^a)^{-1}(\partial_a-i\Phi_a+\frac12
\partial_a\left(\ln\frac{\sqrt{-g}}{H^a}\right))-m)\psi=0
\end{equation}

In a spherically symmetric coordinate system relatively to the
center of the Earth, space-time metric can be chosen in the
following way
\begin{equation}\label{rp2}
ds^2=(1-\frac{r_g}r)dt^2-\frac{dr^2}{1-\frac{r_g}r}-
r^2(\sin^2\theta d\varphi^2+d\theta^2)
\end{equation}
where $r_g=2kM/R$ is Schwarzchild radius.

In a gravitational field the Dirac equation (\ref{urdirr})
relatively to the metric of Schwarzchild (\ref{rp2}) becomes
\cite{Wheeler}
\begin{equation}\label{rp30}
(\gamma^0\frac1f\partial_t+\gamma^r f\partial_r
-\gamma^r\frac{\vec\Sigma\cdot\hat{\vec L}}r-im)e=-\gamma^r(\frac12 f_{,r}+\frac1r(f-1))e,
\end{equation}
where $\vec L=\vec r\times\vec p$ is angular momentum operator,
$f^2=1-\frac{r_g}r$,
$$\gamma^r=\gamma^1\sin\theta\cos\varphi+
\gamma^2\sin\theta\sin\varphi+\gamma^3\cos\theta,\quad
\vec\Sigma=\left(\matrix{\vec\sigma&0\cr0&\vec\sigma}\right).$$

By substituting $f$ in (\ref{rp30}) we find that in the first
approximation
\begin{equation}\label{rp302}
\frac12 f_{,r}+\frac1r(f-1)=-\frac{r_g}{4r^2}
\end{equation}

From (\ref{rp302}) figure out that
\begin{equation}\label{is51}
q^5A_\mu^5=(0,-\frac{r_g}{3r^2},0,0)
\end{equation}

On the other hand, the equation of motion for the charged vector
boson $D_\mu $ is a wave equation for massless particle \cite{Dor2}.
Hence drawing an analogy with an electromagnetic field for which
stationary real component of $1/r $, caused by the energy
conservation law \cite{Xelzen}, is selected, we also will be limited
by the stationary component and we assume that
\begin{equation}\label{is514}
q^5A_\mu^5=q^6A_\mu^6
\end{equation}
or
\begin{equation}\label{is660}
q^5A^5_r=\frac{2kM\hbar}{3c^3r^2}=\frac{2g\hbar}{3c}= 9.8\cdot10^2cm/c^2\cdot10^{-27}erg\cdot c\cdot0.22\cdot10^{-10}c/cm$$
$$=0.2\cdot 10^{-34}erg\approx10^{-23}eV
\end{equation}

\section{Dirac equation in the Riemann-Cartan space}
Covariant derivative of (\ref{kpr}) will be a vector and if the
Christoffel symbols contain an asymmetric part of the lowercase
indices. In this case, pseudoriemann-space is generalized to
Riemann-Cartan or $U^4$ space in which the torsion is equal to the
antisymmetric part of the object
\begin{equation}\label{tk1}
Q^\lambda_{\mu\nu}=\frac12(\Gamma^\lambda_{\mu\nu}-\Gamma^\lambda_{\nu\mu})
\end{equation}
regardless to the pseudo-metric $g_{\mu\nu}$.

Torsion tensor has 24 independent components and can be decomposed
into a sum of three irreducible parts
\begin{equation}\label{tk2}
Q^\lambda_{\mu\nu}=\tilde
Q^\lambda_{\mu\nu}+\frac13(\delta^\lambda_\mu
Q_\nu-\delta^\lambda_\nu  Q_\mu)+ \varepsilon_{\sigma\mu\nu\alpha}
g^{\sigma\lambda}\breve Q^\alpha
\end{equation}
where $\tilde Q^\lambda_{\mu\nu}$ is traceless part of the torsion
tensor, $Q_\mu$ is trace of the torsion tensor
\begin{equation}\label{tks}
Q_\mu=Q^\lambda_{\mu\lambda}
\end{equation}
and $\breve Q^\alpha$ is pseudotrace of the torsion tensor
\begin{equation}\label{tkp}
\breve Q_\alpha=\frac1{3!}\varepsilon_{\alpha\mu\nu\sigma}Q^{\mu\nu\sigma}
\end{equation}
(lifting and lowering indices can be carried out by the metric
tensor $g^{\mu\nu}$).

The part of the object of connectivity, which is symmetric by the
lower-case indexes $\Gamma^\lambda_{\mu\nu}$, is not generally
aligned to the metrics. However further we will limit our
consideration by the case when its symmetric part is defined by a
metric tensor.
\begin{equation}\label{mtkp}
\dot\Gamma^\lambda_{\mu\nu}=\frac12(\Gamma^\lambda_{\mu\nu}+\Gamma^\lambda_{\nu\mu})=\frac12g^{\lambda\sigma}(g_{\nu\sigma,\mu}+g_{\mu\sigma,\nu}-g_{\mu\nu,\sigma})
\end{equation}
hence
\begin{equation}\label{rkp}
\Gamma^\lambda_{\mu\nu}=\dot\Gamma^\lambda_{\mu\nu}+Q^\lambda_{\mu\nu}
\end{equation}

Then the Ricci coefficeint of rotation (\ref{t2}) is generalized to
\begin{equation}\label{t22}
\gamma^b_{ac}=(h^b_{\mu,\nu}+\Gamma^\lambda_{\mu\nu}h^b_\lambda)h^\mu_ah^\nu_c=
(h^b_{\mu,\nu}+\dot\Gamma^\lambda_{\mu\nu}h^b_\lambda +Q^\lambda_{\mu\nu}h^b_\lambda)h^\mu_ah^\nu_c=$$
$$=(h^b_{\mu\tilde;\nu}+Q^\lambda_{\mu\nu}h^b_\lambda)h^\mu_ah^\nu_c
\end{equation}

In some map let's reduce the quadratic form to a diagonal form $M_4$
and we will consider the case with traceless and vector parts of the
tensor of torsion are equal to zero. Then for distinct values
$a,b,c$ in contrast to (\ref{ivsok201}) we will get
\begin{equation}\label{op3}
\Gamma^\lambda_{\mu\nu}h_{c\lambda}h^\mu_bh^\nu_c= Q^\lambda_{\mu\nu}h_{c\lambda}h^\mu_bh^\nu_c\ne0
\end{equation}
and
\begin{equation}\label{op4}
-\gamma^a\Gamma_a=\frac12h^\mu_{c\tilde;\mu}\gamma^c+i\gamma^5\gamma_d(-\frac18\varepsilon^{dabc}h^\mu_ah^\nu_bh_{(c)\lambda} Q^\lambda_{\mu\nu})
\end{equation}
and the Dirac equation takes the form:
\begin{equation}\label{ivsok3}
(\gamma^a(H^a)^{-1}(\partial_a-i\Phi_a+\frac12
\partial_a\left(\ln\frac{\sqrt{-g}}{H^a}\right)+i\gamma^5A_a)-m)\psi=0
\end{equation}
where $A^d=-\frac18\varepsilon^{dabc}h^\mu_ah^\nu_bh_{(c)\lambda} Q^\lambda_{\mu\nu}$.

On the other hand, formerly the Dirac equation on the nonassociative
algebra was obtained in the form (\ref{urdo}). Let's substitute
(\ref{ovd}) in (\ref{urdo}):
\begin{equation}\label{urdor0}
(\gamma^\mu(\partial_\mu+i\frac34q^6A_\mu^6+i\frac34q^6A_\mu^6\gamma^5 +\frac34q^5A_\mu^5+\frac34q^5A_\mu^5\gamma^5)+im)e(x)=0
\end{equation}

In the long derivative the equation has two vectors and two
pseudovectors. Each of these members we consider separately. For
this purpose introduce an index $\epsilon_{g_i}=0,1$, then the long
derivative of the Dirac equation (\ref{urdor0}) becomes
\begin{equation}\label{dpdir}
\partial_\mu+i\epsilon_{g_1}\frac34q^6A_\mu^6+i\epsilon_{g_2}\frac34q^6A_\mu^6\gamma^5+
\epsilon_{g_3}\frac34q^5A_\mu^5 +\epsilon_{g_4}\frac34q^5A_\mu^5\gamma^5
\end{equation}

Assume $\epsilon_{g_1}=1,\epsilon_{g_i}=0,i=2,3,4$. Then we have an
analogy to lengthening of the derivative caused by the
electromagnetic vector-potential
\begin{equation}\label{urdor1}
(\gamma^\mu(\partial_\mu+i\frac34q^6A_\mu^6)+im)e(x)=0
\end{equation}

Therefore we can assume that this term is caused by the interaction
of the electron with the charged pair of $D+D^*$-bosons (as a whole
it is electrically neutral). However since it is several orders
smaller than the conventional electromagnetic interaction of the
electron with surrounding charges of \cite{Dor2}, in general
considered as electrically neutral, that small-order member is to be
neglected.

Assume $\epsilon_{g_3}=1,\epsilon_{g_i}=0,i=1,2,4$. Then we have an
analogy to lengthening of the derivative as a tetrad representation
of the Dirac equation, rewritten in a Riemannian space in a diagonal
metric (\ref{ivsok3})
\begin{equation}\label{urdor3}
(\gamma^\mu(\partial_\mu+\frac34q^5A_\mu^5)+im)e(x)=0
\end{equation}

Assume $\epsilon_{g_2}=1,\epsilon_{g_i}=0,i=1,3,4$. Then we have an
analogy to lengthening of the derivative as a tetrad representation
of the Dirac equation, rewritten in a Riemannian-Cartan space
\begin{equation}\label{urdor2}
(i\gamma^\mu(\partial_\mu+i\gamma^5\frac34q^6A_\mu^6)-m)e(x)=0
\end{equation}

Finally, if $\epsilon_{g_4}=1,\epsilon_{g_i}=0,i=1,2,3$, then in
fact it is the manifestation of the axial properties of the
electromagnetic current (similarly to ST), which is neglected
everywhere below.

\section{Nonrelativistic electron\\ in the space of Riemann-Cartan}
Consider the equation of the electron in the form (\ref{urdor2}),
assuming (\ref{is514}). Let
\begin{equation}\label{is510}
\Psi=e^{-i\varepsilon t/h}\left(\matrix{\varphi(r)\cr\xi(r)}\right)
\end{equation}
then, denoting $T^c=(0,\frac{r_g}{4r^2},0,0)$, obtain
\begin{equation}\label{is60}
\left\{\matrix{\displaystyle\varepsilon\varphi+c\vec\sigma\vec p\xi-mc^2\varphi=-\hbar c\sigma^rT_r^c\varphi\cr\cr\displaystyle
\varepsilon\xi+c\vec\sigma\vec p\varphi+mc^2\xi=-\hbar c\sigma^rT_r^c\xi}\right.
\end{equation}
or, introducing $E'=\varepsilon-mc^2$, have
\begin{equation}\label{is62}
\left\{\matrix{\displaystyle(E'+\hbar c\vec\sigma\vec T^c)\varphi=-c\vec\sigma\vec p\xi\cr\cr\
displaystyle(E'+2mc^2+\hbar c\vec\sigma\vec T^c)\xi=-c\vec\sigma\vec p\varphi}\right.
\end{equation}

If $|E'+T_r^c|<<2mc^2$ then
\begin{equation}\label{is63}
\xi=-(E'+2mc^2+\hbar c\vec\sigma\vec T^c)^{-1}c\vec\sigma\vec p\varphi\approx-\frac1{2mc}\vec\sigma\vec p\varphi
\end{equation}

Substitute (\ref{is63}) in (\ref{is62}). Then for the component
$\varphi$ of the bispinor $\Psi$ we get:
\begin{equation}\label{is630}
E'\varphi\approx(\frac1{2m}(\vec\sigma\vec p)^2-\hbar c\vec\sigma\vec T^c)\varphi
\end{equation}

From the quantum equation (\ref{is630}) it follows that
nonrelativistic particle, with the spin  $\vec\sigma$, in the field
of the Earth has an energy
\begin{equation}\label{en0}
E=\frac{\vec p^2}{2m}+V_g+V_c,
\end{equation}
where $V_g$ is gravitational energy of the particle and $V_c$ is
potential energy of the torsion of Cartan
\begin{equation}\label{env}
V_c=-\hbar c\vec\sigma\vec T^c=-\frac{\hbar r_gc}{4r^3}(\vec\sigma\vec r)
\end{equation}

Thus, in the gravitational field of a massive spherical body the
torsion manifests itself as a shift of spinor particle energy on the
value of
\begin{equation}\label{is65}
\Delta E_c=\frac{\hbar r_gc}{4r^3}(\vec\sigma\vec r)=
\frac {kM\hbar}{2cr^3}(\vec\sigma\vec r)
\end{equation}
were $k=6.67\cdot10^{-8}cm^3g^{-1}c^{-2}$ -- gravitational constant, $M$ -- mass of Earth, $r_g=0.9$ cm
\begin{equation}\label{is66}
\Delta E=\frac{kM\hbar}{2cr^2}=\frac{g\hbar}{2c}= 9.8\cdot10^2cm/c^2\cdot10^{-27}erg\cdot c\cdot0.17\cdot10^{-10}c/cm$$
$$=0.17\cdot 10^{-34}erg\approx10^{-23}eV
\end{equation}

This is sufficiently small value. On the other hand, consider 1
$cm^3$ of the magnetized metal, in which all the spins, causing its
magnetic properties, are oriented in the same direction. Since the
number of spin particles in the volume is equal to  $N=10^{-23}$,
then the energy change due to torsion is
\begin{equation}\label{enk2}
\Delta E_c=N\cdot\Delta E=10^{23}\cdot10^{-23}eV=1 eV
\end{equation}
Now it is more significant. The effect will be even more meaningful
for the ferromagnet, in which a single atom will have several
same-oriented electrons.

Another estimation of the impact of torsion to the satellite is
interesting. Let the satellite be an oriented permanent magnet. Its
energy is
\begin{equation}\label{enm2}
E=\frac{mv^2}2-k\frac{Mm}r-\frac{N\hbar
r_g(\hat{\vec\sigma}\cdot\vec r)}{4cr^3}
\end{equation}
where $N$ is a number of the spins.

Suppose there is a rotation around a circular orbit with a total
energy equal to zero. Then $v=\Omega r$ and
\begin{equation}\label{enm3}
\frac m2\Omega^2r^2=\frac{kMm}r+\frac{N\hbar
kM(\hat{\vec\sigma}\cdot\vec r)}{2cr^3}
\end{equation}

Hence
\begin{equation}\label{enm4}
\frac{\Omega^2_1}{\Omega^2_2}=\frac{r^3_2}{r^3_1}\left(1+\frac{N\hbar}{mc}
\left(\frac{(\hat{\vec\sigma}\cdot\vec
r_1)}{r^2_1}-\frac{(\hat{\vec\sigma}\cdot\vec
r_2)}{r^2_2}\right)\right)
\end{equation}
where $(\hat{\vec\sigma}\cdot\vec r)=\pm\frac12r$. If the directions
of $\hat{\vec\sigma}$ and $\vec r$ match to each other then
$(\hat{\vec\sigma}\cdot\vec r)=\frac12r$. If the directions are
opposite, then $(\hat{\vec\sigma}\cdot\vec r)=-\frac12r$. We
estimate the correction to Kepler's third law, which follows from
(\ref{enm4}) for the satellite, in which every atom has an oriented
spin (then $N/m=1/m_A$), with $m_A$ standing by atomic mass. Let one
 oriented spin accounts for ten protons, then
\begin{equation}\label{enm5}
\frac\hbar{m_Acr}=10^{-23}
\end{equation}

\section{Conclusion}
The energy of (\ref{is66}), per one spin, is too small for modern
physical effects. Even one mole of a substance gives the
contribution of only one electron-volt. However research of the
interaction of the matter spin and gravity would deepen our
knowledge about the surrounding macrocosm, and perhaps a microcosm.
The effects of torsion could play a role in the case of cosmic
objects. Particularly, the disk structure of the space jet induces a
torsion perpendicular to the plane of the jet, which might be a
cause of observed polarization in the jets and jet emission
currently assigned to the influence of the magnetic field jet.

There is another effect associated with this torsion. If you watch a
satellite, moving by a parallelogram, then the parallelogram does
not close and the defect is \cite{Obuhov}:
\begin{equation}\label{z1}
\Delta x^\alpha=\Gamma^\alpha_{\beta\gamma}S^{\beta\gamma}
\end{equation}
where $S^{\beta\gamma}$ is a pseudotensor of the parallelogram
square. Since the pseudotensor of the square is antisymmetric by the
indices and as it is clear from (\ref{z1}), the effect is entirely
given by torsion, particularly by its traceless part. Taking the
circle instead of the parallelogram, in the case of the torsion
(\ref{ivsok3}) we obtain
\begin{equation}\label{z1}
\Delta x=\pi R^2\cdot\frac{r_g}{3R^2}\approx r_g\approx1 cm
\end{equation}

This value gives the angular dimesion which is observed from the
Earth while watching to the satellite from the distance of 100 km:
\begin{equation}\label{z2}
\Delta\varphi=\frac{\Delta x}R=\frac{1 cm}{100 km}\approx10^{-7},
\end{equation}
which corresponds to the one thousandth of the second per one
revolution.


\begin{thebibliography}{99}
\bibitem{Dor1}
Dorofeev V. Yu. arxiv.org math-ph:0908.3247v1 (2009)
\bibitem{Dor2}
Dorofeev V. Yu. arxiv.org gr-qc:1003.3228 (2010)
\bibitem{Nest}
Nesterov A. I. and Sabinin L. V. Phys. Rev. D62, 081501 (2000).
\bibitem{Gob}
Gogberashvili M. hep-th/0409173.
\bibitem{FokIvanenko}
Fock V., Ivanenko D., C. R., Paris 188, 1470 (1929).
\bibitem{Eisenhart}
Eisenhart L. P. \emph{Riemannian gyometry.} 1926.
\bibitem{SokolovIvanenko}
Sokolov A. Ivanenko D. \emph{Quantum theory field.} 1952.
\bibitem{Xelzen}
Helzen F., Martin A. \emph{Quarks and leptons.} 1987.
\bibitem{Wheeler}
Brill D. R., Wheeler J. A. Rev. Mod. Phys., {\bf 29}, 465, (1957).
\bibitem{Okun}
Okun L. B. {\it Leptons and quarqs.} 1990.
\bibitem{Obuhov}
Barvinsky A. O., Obuhov Yu. N., Ponomaryev V. N. {\it Of the geometrydynamic methods and the gage approach to the theory of gravitational interactions..} 1985.
\end{thebibliography}
\end{document}